\documentclass[aip,pop,reprint]{revtex4-1}
\pdfoutput=0
\usepackage{amsmath}
\usepackage{amssymb}

\usepackage[utf8]{inputenc}
\usepackage[T1]{fontenc}
\usepackage{lmodern,stmaryrd,xcolor,graphicx}
\usepackage[francais,english]{babel}

\begin{document}

\title{Hamiltonian fluid closures of the Vlasov-Amp\`ere equations: from water-bags to $N$ moment models}

\author{M. Perin}

\author{C. Chandre}

\affiliation{Aix-Marseille Universit\'e, Universit\'e de Toulon, CNRS, CPT UMR 7332, 13288 Marseille, France}

\author{P.J. Morrison}

\affiliation{Department of Physics and Institute for Fusion Studies, The University of Texas at Austin, Austin, TX 78712-1060, USA}

\author{E. Tassi}

\affiliation{Aix-Marseille Universit\'e, Universit\'e de Toulon, CNRS, CPT UMR 7332, 13288 Marseille, France}

\begin{abstract}
Moment closures of the Vlasov-Amp\`ere system, whereby higher moments are represented as functions of lower moments with the constraint that the resulting fluid system remains Hamiltonian, are investigated by using  water-bag theory.  The link between the water-bag formalism and fluid models that involve  density, fluid velocity,  pressure and higher moments is established by introducing suitable thermodynamic variables. The cases of one, two and three water-bags are treated and their Hamiltonian structures are provided. In each case, we give the associated fluid closures and we discuss their Casimir invariants. We show how the method can be extended to an arbitrary number of fields, i.e., an arbitrary number of water-bags and associated moments. The thermodynamic interpretation of the resulting models is discussed. Finally, a general procedure to derive Hamiltonian $N$-field fluid models is proposed.
\end{abstract}

\maketitle

\section{Introduction}

Due to their high temperature, many plasmas, such as the ones encountered in the core of tokamaks or in the magnetosphere, can be considered as collisionless.  Consequently, they may be well described by kinetic equations such as the Vlasov-Maxwell system where the particle dynamics is described by  a distribution function  $f(\mathbf{x},\mathbf{v},t)$  defined on a six-dimensional phase space $(\mathbf{x},\mathbf{v})$ with $\mathbf{x}$  being  the position and $\mathbf{v}$ the velocity. It is particularly challenging to solve such kinetic models, even using advanced numerical techniques. In addition, a full kinetic description of the system might provide unnecessary  information,  depending on the phenomena under investigation. This is the justification, e.g., for gyrokinetic theories where a strong magnetic field assumption\cite{Brizard07} leads to the perpendicular component of the velocity  being replaced by the magnetic moment considered as an adiabatic invariant.  Similarly,  anisotropy due to a strong  magnetic field  is also used to reduce the complexity of the original kinetic problem in  the double adiabatic theory\cite{Chew56}.  More generally, and ideally, reduced models obtained by some kind of closure leads  to a decrease in  the complexity of the original  kinetic problem,  while maintaining accuracy and providing  physical insights about the processes at work.  Consequently, fluid reductions of kinetic equations are often sought.

Generally speaking,  fluid models are obtained by projecting the distribution function as follows:
\begin{equation}
\label{introF}
f(\mathbf{x},\mathbf{v},t)\simeq\sum\limits_{i=0}^NP_i(\mathbf{x},t)e_i(\mathbf{v}),
\end{equation}
where $P_i(\mathbf{x},t)$ is the $i$-th fluid moment defined as the $i$-th moment of the distribution function with respect to the velocity $\mathbf{v}$, i.e., 
\begin{equation*}
P_i(\mathbf{x},t)=\int\underset{i\text{ times}}{\underbrace{\mathbf{v}\otimes\mathbf{v}\dots\otimes\mathbf{v}}}f(\mathbf{x},\mathbf{v},t)\ \mathrm{d}\mathbf{v},
\end{equation*}
for all $i\in\mathbb{N}$ and $\{e_i(\mathbf{v})\}$ is some specific fixed set of basis functions. The dynamics of the distribution function $f(\mathbf{x},\mathbf{v},t)$ is then given by the dynamics of the fluid moments that  are functions of  the configuration space coordinate $\mathbf{x}$ only. This makes fluid models, which involve quantities such as, the density $\rho(\mathbf{x},t)=P_0(\mathbf{x},t)$, the fluid velocity $\mathbf{u}(\mathbf{x},t)=P_1(\mathbf{x},t)/P_0(\mathbf{x},t)$ and the pressure $P(\mathbf{x},t)=P_2(\mathbf{x},t)-P_1^2(\mathbf{x},t)/P_0(\mathbf{x},t)$,   convenient to interpret. Furthermore, since fluid variables  only depend on $\mathbf{x}$ at each time, they are substantially less expensive to solve numerically than their kinetic counterpart.  

Clearly accurate reduced fluid models are desirable, but finding effective fluid closures is a difficult and largely open problem. Indeed, despite their strong physical relevance,  a general or optimal procedure for obtaining them for  the Vlasov equation  does not exist.    For example, consider the following simple free advection equation:
\begin{equation}
\label{eqAdvection}
\partial_t f=-\mathbf{v}\cdot\nabla f\,,
\end{equation}
which is the Vlasov equation with field dynamics removed. Multiplying Eq.~\eqref{eqAdvection} by $\mathbf{v}^n$ and integrating with respect to the velocity, yields the following infinite hierarchy  of moment equations:
\begin{equation}
\label{eqAdvectionP}
\partial_tP_i=-\nabla\cdot P_{i+1},
\end{equation}
for all $i\in\mathbb{N}$. In order to be able to solve Eqs.~\eqref{eqAdvectionP}, one has to truncate the infinite set of equations at some order $N\in\mathbb{N}$. However, because the time evolution of $P_N$ involves $P_{N+1}$, the latter must  be neglected or expressed in terms of lower order moments, i.e.,  $P_{N+1}=P_{N+1}(P_{i\leq N})$. This is the ubiquitous closure problem for  fluid reductions of kinetic equations.

In conventional fluid closure theory a  collision process and assumption of local thermodynamic equilibrium is basic. Instead, in this article we consider  Hamiltonian fluid reductions, where we investigate closures based on whether or not they preserve Hamiltonian structure.  This  allows us to select out a subset of all possible  fluid closures that preserve  the geometrical structure and this  prevents the introduction of non-physical dissipation in the resulting fluid moment system, without requiring nearness to thermal equilibrium. 

The  case of two moments\cite{Gibbons81}  corresponds to the well-known exact  water-bag reduction, so  it is not surprising that it is Hamiltonian.   However, as one increases the number of moments, this increases  the dimension of the subset $P_{i\leq N}$, and the constraints needed to preserve the Hamiltonian  structure become more difficult to solve\cite{Perin14,Perin15}, so that  eventually, it is not possible to obtain a general  analytic expression for the closure $P_{N+1}=P_{N+1}(P_{i\leq N})$.

The problem of deriving Hamiltonian fluid models can, however, be tackled from  different angles. Indeed, instead of Eq.~\eqref{introF}, other representations of the distribution function can be used to decrease the complexity of the initial problem. In this paper we consider a general  water-bag model\cite{Roberts67,Bertrand68,Berk70}, which has  also been used, e.g., in gyrokinetics\citep{Morel07,Gravier08,Morel08,Gravier13}. In one dimension, this projection is obtained by replacing the distribution function with a piecewise constant function in the velocity $v$ such that
\begin{equation}
\label{introWB}
f(x,v,t)\simeq\sum\limits_{i=1}^{N+1}a_i\Theta[v-\mathrm{v}_i(x,t)],
\end{equation}
where $a_i$ are constants, $\Theta$ denotes the Heavyside distribution and $\mathrm{v}_i(x,t)$ is a set of contour velocities. Like with the fluid moment projection, the dynamics of the distribution function $f(x,v,t)$ defined on  phase space has been replaced by the dynamics of $N+1$ fields defined on  configuration space, namely $\mathrm{v}_i(x,t)$ for all $1\leq i\leq N+1$.

The use of the water-bag projection constitutes an exact reduction and consequently  the resulting system is intrinsically Hamiltonian\citep{Yu00,Chesnokov12}. When the number of field variables expressed in terms of  fluid moments and the number of contour velocities are the same,  the water-bag projection is easier to handle than that of the usual fluid moments representation; in particular, this is the case for  the computation of a Poisson bracket.  However, even though the contour velocities $\mathrm{v}_i$ are rather convenient to handle from a computational point of view, their macroscopic physical interpretation is less obvious than for the fluid moments. Consequently, there is a balance to seek between the computational simplicity of the closure provided by the water-bag model and the physical relevance of the fluid moments.

In this article, we investigate links between the water-bag and the fluid moment representations in order to generate new Hamiltonian closures. Indeed, any truncation of the infinite series given by Eq.~\eqref{introWB} is preserved by the dynamics and hence constitutes a closure. As a consequence, the subset of all the water-bag distribution functions is invariant. Following the water-bag projection, we perform a fluid reduction of the distribution function to obtain a Hamiltonian fluid model. Then, we construct a systematic procedure to obtain a fluid reduction from the water-bag distribution function by preserving the Hamiltonian structure of the parent kinetic model. We extend this procedure to build general $N$-field Hamiltonian fluid models with $N-2$ internal degrees of freedom.

In Sec.~\ref{secVA}, we provide the Hamiltonian structure of the Vlasov-Amp\`ere equations which constitute the parent kinetic model. The Casimir invariants of the associated bracket are provided. We introduce the water-bag distribution function and give the associated Hamiltonian structure. Some properties of the system such as invariants are discussed. In Sec.~\ref{secNewVar}, we establish a link between the water-bag and the fluid models. This is done by exhibiting a peculiar set of fluid variables that allows us to make explicit the fluid closure corresponding to the water-bag model. We use the density and the fluid velocity to account for the macroscopic energy of the system and we propose suitable variables to take into account internal degrees of freedom coming from microscopic phenomena. The Hamiltonian structure of the resulting equations is provided and their Casimir invariants are discussed. We also address the thermodynamic implications of the new variables. Lastly in Sec.~\ref{secNWB}, new models are proposed to extend the results obtained from the water-bag model to more general distribution functions. This allows us to construct general $N$-field fluid models that describe plasmas with $N-2$ internal degrees of freedom.

\section{The Hamiltonian structure of the Vlasov-Amp\`ere equations and the water-bag model}
\label{secVA}

We investigate the dynamics of a one-dimensional plasma made of electrons of unit mass and negative unit charge evolving in a background of static ions.  This simplified system contains essential difficulties of more complete dynamics.  We assume vanishing boundary conditions at infinity in velocity $v$ and periodic boundary conditions in the spatial domain of unit length. The time evolution of the distribution function of the electrons $f(x,v,t)$ and the electric field $E(x,t)$ is described by the one-dimensional Vlasov-Amp\`ere equations, 
\begin{align}
\label{eqVlasov}
\partial_t f&=-v\partial_x f+\widetilde{E}\partial_v f,\\
\label{eqAmpere}
\partial_t E&=-\widetilde{\jmath},
\end{align}
where $\widetilde{E}=E-\int E\ \mathrm{d}x$ and $\widetilde{\jmath}=j-\int j\ \mathrm{d}x$ are the fluctuating parts of the electric field $E(x,t)$ and the current density $j(x,t)=-\int vf(x,v,t)\ \mathrm{d}v$,  respectively.

The Vlasov-Amp\`ere model possesses a Hamiltonian structure for the distribution function of the electrons $f(x,v,t)$ and the electric field $E(x,t)$. The Poisson bracket acting on functionals $F[f,E]$ is\citep{Chandre13,Morrison82,Morrison98}
\begin{multline}
\label{eqBrackVA}
\{F,G\}=\int f\big[\partial_x F_f\partial_v G_f-\partial_x G_f\partial_v F_f\\
+\widetilde{F_E}\partial_v G_f-\widetilde{G_E}\partial_v F_f\big]\ \mathrm{d}x\mathrm{d}v,
\end{multline}
where $F_f$  and  $F_E$ denote the functional derivative of $F$ with respect to the distribution function $f(x,v,t)$ and  the electric field $E(x,t)$, respectively. Bracket~\eqref{eqBrackVA} is a Poisson bracket, i.e., it satisfies four essential properties: it is linear in both its arguments; it is alternating, i.e., $\{F,F\}=0$; it satisfies the Leibniz rule, i.e., $\{F,GH\}=\{F,G\}H+G\{F,H\}$; it verifies the Jacobi identity, i.e.,
\begin{equation*}
\{F,\{G,H\}\}+\{H,\{F,G\}\}+\{G,\{H,F\}\}=0.
\end{equation*}
for all functionals $F$, $G$ and $H$. The Hamiltonian of the system, which corresponds to the total energy, is 
\begin{equation}
\label{hamVA}
\mathcal{H}[f,E]=\int\frac{v^2}{2}f\ \mathrm{d}x\mathrm{d}v+\int\frac{E^2}{2}\ \mathrm{d}x,
\end{equation}
where the first term accounts for the kinetic energy of the electrons and the second term corresponds to the energy of the electric field. Bracket~\eqref{eqBrackVA} and Hamiltonian~\eqref{hamVA} lead to Eqs.~\eqref{eqVlasov} and \eqref{eqAmpere} by using $\partial_t f=\{f,\mathcal{H}\}$ and $\partial_t E=\{E,\mathcal{H}\}$.

The Casimir invariants of a bracket $\{\cdot,\cdot\}$ are particular observables $C$ that commute with all observable $F$, i.e.,  $\{F,C\}=0$ for all functionals $F$. Bracket~\eqref{eqBrackVA} possesses a local ($x$-dependent) Casimir invariant given by
\begin{equation}
\label{Gauss}
C_\text{loc}=\partial_xE+\int f\ \mathrm{d}v,
\end{equation}
which corresponds to Gauss's law. There are also  global ($x$-independent) invariants.  Namely, 
\begin{equation}
\label{meanE}
\bar{E}=\int E\ \mathrm{d}x,
\end{equation}
which expresses the fact that the mean value of the electric field remains constant. This results from the periodic boundary conditions in space and from the definition of the electric field $E=-\partial_x\Phi$ where $\Phi(x,t)$ is the electrostatic potential. Finally, there is a family of global invariants given by
\begin{equation}
\label{casPhi}
C_1=\int\phi(f)\ \mathrm{d}x\mathrm{d}v,
\end{equation}
where $\phi(f)$ is any function of $f$. This family of Casimir invariants arises from particle  relabeling symmetry and includes, e.g.,  the cases of conservation of the total mass and the usual entropy.

\begin{figure}
\centering
\includegraphics[width=0.4\textwidth]{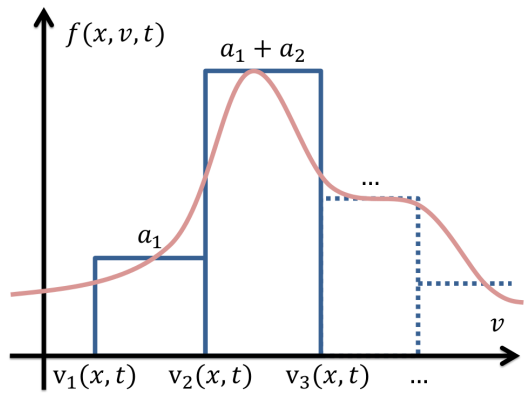}
\caption{Sketch of a distribution function (in light red) and its water-bag approximation (in dark blue).}
\label{figWaterBags}
\end{figure}

The water-bag model is a particular solution of Eqs.~\eqref{eqVlasov} and \eqref{eqAmpere} with a piecewise constant initial condition for the distribution function $f(x,v,t)$, 
\begin{equation}
\label{defWaterBag}
f_N(x,v,t)=\sum\limits_{i=1}^{N+1}a_i\, \Theta[v-\mathrm{v}_i(x,t)],
\end{equation}
which can be done for any  $N\in\mathbb{N}$.   In water-bag theory one is interested in approximating a smooth initial condition by a water-bag approximation,  such as that shown in Fig.~\ref{figWaterBags}.
In order for this distribution function to have  compact support, we further require the following constraint:
\begin{equation*}
\sum\limits_{i=1}^{N+1}a_i=0.
\end{equation*}
Moreover, the velocities $\mathrm{v}_i(x,t)$ are supposed to be ordered such that for all $(x,t)\in[0;1[\times\mathbb{R}_+$ we have $\mathrm{v}_1(x,t)<\mathrm{v}_2(x,t)<\dots<\mathrm{v}_{N+1}(x,t)$. In what follows, we will refer to this distribution function as an $N$-water-bag distribution function. A distribution function of the  form of  Eq.~\eqref{defWaterBag} is solution of Eqs.~\eqref{eqVlasov} and \eqref{eqAmpere}, and hence its  form is preserved by the dynamics, if and only if the contour velocities $\mathrm{v}_i(x,t)$ satisfy \citep{Roberts67,Bertrand68,Berk70,Morel07,Gravier08,Morel08,Besse09,Gravier13}
\begin{equation}
\label{eqContours}
\partial_t\mathrm{v}_i=-\mathrm{v}_i\partial_x\mathrm{v}_i-\widetilde{E},
\end{equation}
for all $1\leq i\leq N+1$ and
\begin{equation}
\label{eqAmpereContours}
\partial_t E=-\frac{1}{2}\sum\limits_{i=1}^{N+1}a_i\widetilde{\mathrm{v}_i^2}.
\end{equation}
As a consequence, solving the Vlasov equation for a water-bag distribution function $f_N(x,v,t)$ is equivalent to solving the $N+1$ contour equations given by Eq.~\eqref{eqContours} for the contour velocities $\mathrm{v}_i(x,t)$. Coupling between the different contours is then provided by Eq.~\eqref{eqAmpereContours}\cite{foot1}.

The water-bag model possesses a Hamiltonian structure\cite{Chesnokov12,Morrison14a,Morrison14b} inherited from the original Vlasov-Amp\`ere equations. This means that there exist a bracket $\{,\}_{WB}$ and a Hamiltonian $\mathcal{H}$ such that Eqs.~\eqref{eqContours}-\eqref{eqAmpereContours} are obtained by $\partial_t \mathrm{v}_i=\{\mathrm{v}_i,\mathcal{H}\}_{WB}$ for all $1\leq i\leq N+1$ and $\partial_t E=\{E,\mathcal{H}\}_{WB}$,  respectively. For the water-bag distribution function of  Eq.~\eqref{defWaterBag}, the dynamical variables are the contour velocities $\mathrm{v}_i(x,t)$ and the electric field $E(x,t)$. Using the chain rule, the functional derivative of $F$ with respect to $\mathrm{v}_i$, denoted $F_i$, is 
\begin{equation}
\label{fctDerivative}
F_i=-a_iF_f\big|_{v=\mathrm{v}_i},
\end{equation}
for all $1\leq i\leq N+1$. Along with Eqs.~\eqref{defWaterBag} and  Eq.~\eqref{fctDerivative} this leads to the water-bag bracket\cite{Morrison14a,Morrison14b}
\begin{equation}
\label{eqBrackWB}
\{F,G\}_{WB}=\sum\limits_{i=1}^{N+1}\int\bigg[\frac{1}{a_i}F_i\partial_x G_i+G_i\widetilde{F_E}-F_i\widetilde{G_E}\bigg]\ \mathrm{d}x.
\end{equation}
One can show that Bracket~\eqref{eqBrackWB} is a Poisson bracket, which is a property inherited from the Vlasov-Amp\`ere equations. The Hamiltonian of the system is \begin{equation}
\label{hamWB}
\mathcal{H}[\mathrm{v}_1,\dots,\mathrm{v}_{N+1},E]=\frac{1}{2}\int\left[-\frac{1}{3}\sum\limits_{i=1}^{N+1}a_i\mathrm{v}_i^3+E^2\right]\ \mathrm{d}x\,, 
\end{equation}
which is obtained from Eq.~\eqref{hamVA} by using Eq.~\eqref{defWaterBag}. 

An important feature of Bracket~\eqref{eqBrackWB} is that it is closed. Thus, for any number of bags $N\in\mathbb{N}$, the set of the water-bag distribution functions $f_N$ of $N$ bags is a sub-Poisson algebra of the Vlasov-Amp\`ere model. This means that the Vlasov-Amp\`ere dynamics preserves the number of bags. In particular, the water-bag model is Hamiltonian for any  number of bags. This is particularly interesting from a numerical point of view, e.g., as there is no nonphysical dissipation introduced by the water-bag approximation even for a low order approximation with a small number of bags.

Bracket~(\ref{eqBrackWB}) possesses several Casimir invariants. By using Eq.~\eqref{defWaterBag}, Eq.~\eqref{Gauss} becomes
\begin{equation*}
C_\text{loc}=\partial_xE-\sum\limits_{i=1}^{N+1} a_i\mathrm{v}_i.
\end{equation*}
The global invariant given by Eq.~\eqref{meanE} is preserved by Eq.~(\ref{eqBrackVA}). The family of Casimir invariants given by Eq.~\eqref{casPhi} is projected to 
\begin{equation*}
C_1=\sum\limits_{i=1}^N\phi\left(A_i\right)(\bar{\mathrm{v}}_{i+1}-\bar{\mathrm{v}}_i),
\end{equation*}
where $A_i=\sum\limits_{k=1}^ia_k$ and 
\begin{equation*}
\bar{\mathrm{v}}_i=\int\mathrm{v}_i(x,t)\ \mathrm{d}x,
\end{equation*}
for all $1\leq i\leq N+1$. As $C_1$ is a Casimir invariant for any function $\phi$, this shows that the projection of the invariant given by Eq.~\eqref{casPhi} leads to the generation of $N$ invariants, namely
\begin{equation*}
C_{1,i}=\bar{\mathrm{v}}_{i+1}-\bar{\mathrm{v}}_i,
\end{equation*}
for all $1\leq i\leq N$. However, $C_1$ is computed such that $[f,\delta C_1/\delta f]=0$ for any distribution function $f$, where $[g,h]=\partial_xg\partial_vh-\partial_xh\partial_vg$. If we now look only at water-bag distribution functions, the requirement for $C_1$ becomes $[f,\delta C_1/\delta f]=0$ for all $f$ given by Eq.~\eqref{defWaterBag}, and hence is less restrictive. This leads to the creation of an additional invariant, e.g., $\bar{\mathrm{v}}_1$. Thus,  there are $N+1$ Casimir invariants given by $\bar{\mathrm{v}}_i$ for all $1\leq i\leq N+1$, i.e., as many Casimir invariants as the number of fields.\cite{foot2}

\section{Link between the water-bag model and the fluid moments of the distribution function}
\label{secNewVar}

The contour velocities $\mathrm{v}_i(x,t)$ provide immediate kinetic theory  information:  they define the partitioning of the distribution function in the velocity space, sorting particles into water-bags according to  their velocities.  However, their interpretation on the fluid level in terms of moments $P_i(x,t)$,   given by
\begin{equation}
\label{defP}
P_i(x,t)=\int v^if(x,v,t)\ \mathrm{d}v 
\end{equation}
for all $i\in\mathbb{N}$,  is not so clear.  This relationship is given explicitly by inserting  Eq.~\eqref{defWaterBag} into  Eq.~\eqref{defP}, yielding 
\begin{equation}
\label{eqPWB}
P_i(x,t)=\frac{-1}{i+1}\sum\limits_{k=1}^{N+1}a_k\mathrm{v}_k^{i+1}(x,t),
\end{equation}
for all $i\in\mathbb{N}$. Our strategy  is to use the information provided by the water-bag model to build Hamiltonian models for the fluid moments $P_i(x,t)$. Indeed, as stated in Sec.~\ref{secVA}, the water-bag model possesses a Hamiltonian structure. As a consequence, by expressing the contour velocities with respect to the fluid moments we can obtain particular fluid models with an arbitrary number of moments. We believe that this is a useful strategy because constructing general fluid models can be technically very challenging\cite{Perin15}. The closures provided by the water-bag models provide  insight for  building more general  fluid models.

\subsection{The single water-bag model}
\label{secSingle}

First consider the  case of a single water-bag in order to illustrate our approach,  which will be generalized to an arbitrary number of water-bags corresponding to  an arbitrary number of fluid moments. This simple model constitutes a good illustration of our strategy. If we consider a single water-bag, or equivalently two contour velocities, the distribution function simply reads
\begin{equation*}
f_1(x,v,t)=\Theta[v-\mathrm{v}_1(x,t)]-\Theta[v-\mathrm{v}_2(x,t)].
\end{equation*}
where we have set $a_1=1$ defining  unit height  to the  water-bag. 
Using Eq.~\eqref{eqPWB}, the first two moments of the distribution function are 
\begin{equation*}
P_0=\mathrm{v}_2-\mathrm{v}_1\qquad \mathrm{and} \qquad P_1=\frac{\mathrm{v}_2^2-\mathrm{v}_1^2}{2}.
\end{equation*}
which upon inversion yield
\begin{equation*}
\mathrm{v}_1=\frac{P_1}{P_0}-\frac{P_0}{2}  \qquad \mathrm{and} \qquad  \mathrm{v}_2=\frac{P_1}{P_0}+\frac{P_0}{2}.
\end{equation*}
Defining the density $\rho=P_0$ and the fluid velocity $u=P_1/P_0$, we obtain the fluid variables 
\begin{equation*}
\mathrm{v}_1=u-\frac{\rho}{2} \qquad \mathrm{and} \qquad \mathrm{v}_2=u+\frac{\rho}{2}.
\end{equation*}
As a consequence, we are able to express the contour velocities with respect to the usual fluid moments. In terms of these fluid variables, the water-bag bracket given by Eq.~\eqref{eqBrackWB} is 
\begin{equation*}
\{F,G\}_1=\int\big[G_u\partial_x F_\rho-F_u\partial_xG_\rho+G_u\widetilde{F_E}-F_u\widetilde{G_E}\big]\ \mathrm{d}x,
\end{equation*}
where now $F_\rho$ and  $F_u$ denote the functional derivative of $F$ with respect to $\rho$ and  $u$, respectively. This bracket, which corresponds to the cold-plasma bracket\cite{Morrison98}, is closed, a  property  inherited from the original water-bag bracket given by Eq.~\eqref{eqBrackWB}.  In terms of the variables $\rho$, $u$ and $E$, Hamiltonian~\eqref{hamWB} becomes
\begin{equation*}
\mathcal{H}[\rho,u,E]=\int\left(\frac{1}{2}\rho u^2+\rho U(\rho)+\frac{E^2}{2}\right)\ \mathrm{d}x,
\end{equation*}
where $U(\rho)=\rho^2/24$ is the specific internal energy of the system. The pressure is defined by the usual thermodynamic relation  $P(\rho)=\rho^2\partial U/\partial\rho=\rho^3/12$. The reduced moments, defined by
\begin{equation}
\label{eqS}
S_i(x,t)=\frac{1}{\rho^{i+1}}\int (v-u)^if(x,v,t)\ \mathrm{d}v,
\end{equation}
for all $i\geq 2$, appear to be suitable variables to describe the Poisson structure of the fluid equations resulting from the Vlasov-Amp\`ere model\cite{Perin15}. With this definition, the second order reduced moment reads $S_2=P/\rho^3$,  which eventually leads to $S_2=1/12$, i.e., $S_2$ is constant. (Note, throughout we will express the $S_i$ in $1/a_1^i$ units for all $i\geq2$.)

In term of the reduced moments, the Hamiltonian of the system reads
\begin{equation*}
\mathcal{H}[\rho,u,E]=\frac{1}{2}\int\left(\rho u^2+\rho^3S_2+E^2\right)\ \mathrm{d}x\,, 
\end{equation*}
and the equations of motion are
\begin{align}
\label{eqContinuity}
\partial_t\rho&=\{\rho,\mathcal{H}\}_1=-\partial_x(\rho u),\\
\label{eqEuler}
\partial_tu&=\{u,\mathcal{H}\}_1=-u\partial_xu-\frac{1}{\rho}\partial_x(\rho^3S_2)-\widetilde{E},\\
\label{eqAmpereRhoU}
\partial_tE&=\{E,\mathcal{H}\}_1=\widetilde{\rho u}.
\end{align}
These are the equations for a barotropic fluid undergoing an adiabatic process. Indeed, the relationship between the pressure $P$ and the density $\rho$ is such that $P/\rho^3=S_2=1/12$ is a constant. This is the characteristic isentropic equation of state of an ideal gas with one degree of freedom. The local Casimir invariant given by Eq.~\eqref{Gauss} is preserved and is 
\begin{equation}
\label{Gauss2}
C_\text{loc}=\partial_xE+\rho.
\end{equation}
In addition, the system have three global invariants: the mean value of the electric field given by Eq.~\eqref{meanE} and
\begin{align}
\label{casRho}
\bar{\rho}&=\int\rho\ \mathrm{d}x,\\
\bar{u}&=\int u\ \mathrm{d}x.
\end{align}
The first invariant is  conservation of  total mass,  which results from the fact that the system is isolated. This Casimir invariant results from the projection of Eq.~\eqref{casPhi}. As for the electric field, the last Casimir invariant, corresponds to  conservation of the mean value of the velocity. As noted, this additional invariant arises from the closure procedure.

\subsection{Two water-bag  model: introduction of the thermodynamical variables}
\label{secDouble}

\begin{figure}
\centering
\includegraphics[width=0.4\textwidth]{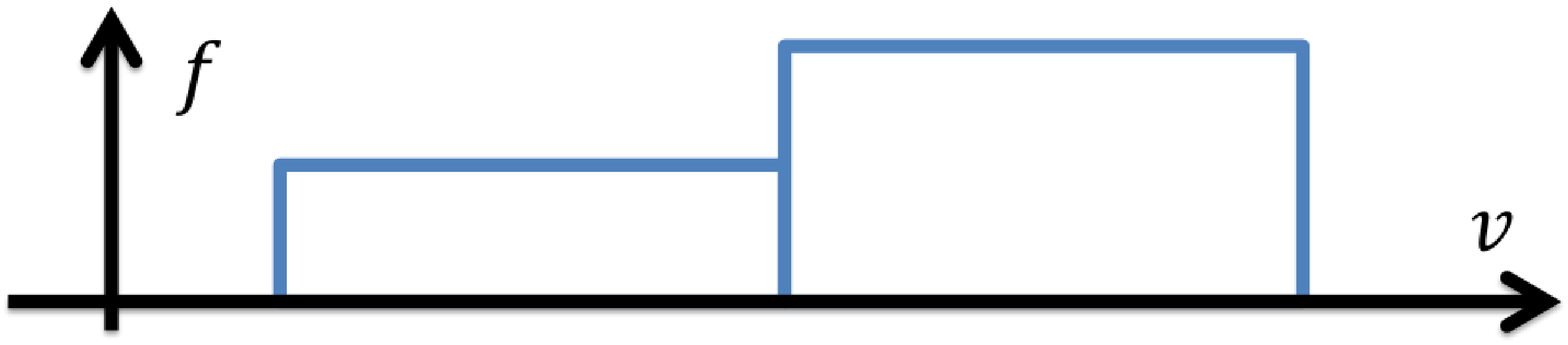}
\includegraphics[width=0.4\textwidth]{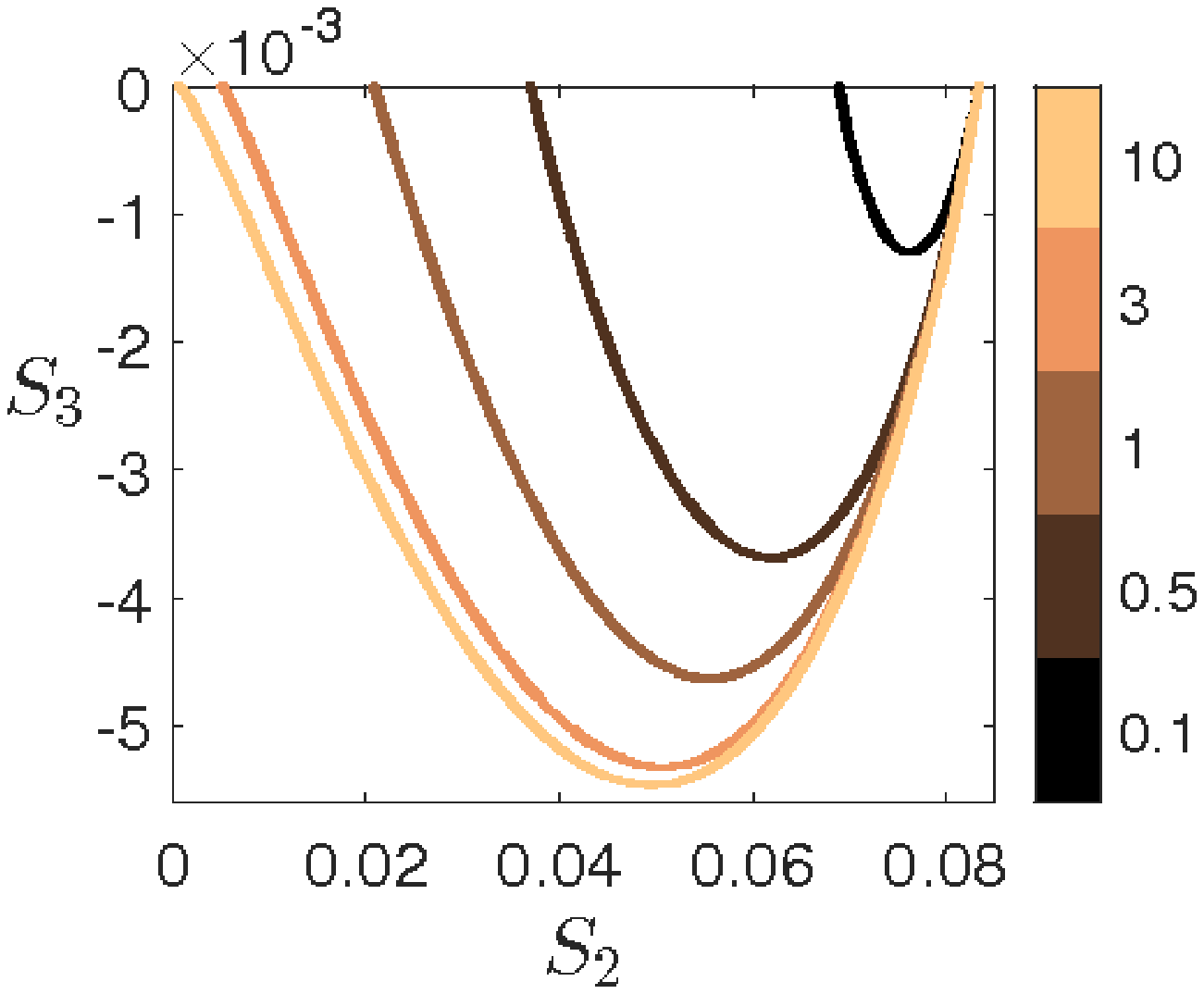}
\caption{Upper panel: sketch of a double water-bag distribution function. Lower panel: plot of $S_3$ as a function of $S_2$ for a double water-bag distribution function corresponding to the upper panel and for different values of $a_2$ (given by the colorbar).}
\label{figS3}
\end{figure}

Now consider the  case of two water-bags,    in order to characterize more precisely the closure provided by the water-bag model and its relation to the fluid moments. Indeed, there exists an infinite number of Hamiltonian fluid models with three moments\cite{Perin14}. By using the reduced moments defined by Eq.~\eqref{eqS}, Hamiltonian models for the variables $\rho$, $u$ and $S_2$ are such that $S_3$ is an arbitrary function of $S_2$.  Two water-bags means we have  three contour velocities or, equivalently, three fluid moments, which provides a particular example of the more general closure $S_3=S_3(S_2)$. Such a distribution function, whose expression is given by Eq.~\eqref{defWaterBag} with $N=2$, is represented in Fig.~\ref{figS3}. By using Eq.~\eqref{eqPWB}, the first three moments of the distribution function are 
\begin{align*}
P_0&=\mathrm{v}_3-\mathrm{v}_1+a_2(\mathrm{v}_3-\mathrm{v}_2),\\
P_1&=\frac{\mathrm{v}_3^2-\mathrm{v}_1^2}{2}+a_2\frac{\mathrm{v}_3^2-\mathrm{v}_2^2}{2},\\
P_2&=\frac{\mathrm{v}_3^3-\mathrm{v}_1^3}{3}+a_2\frac{\mathrm{v}_3^3-\mathrm{v}_2^3}{3}.
\end{align*}
Unlike the single water-bag case, expressing the contour velocities with respect to the fluid moments with two water-bags is more complicated. This is partially due to the fact that $P_2$ involves cubic terms in the contour velocities $\mathrm{v}_i$. This issue becomes more acute as the number of water-bags increases. A first step toward the fluid representation can be easily done by using the following variable:
\begin{equation*}
n_1=\frac{\mathrm{v}_{2}-\mathrm{v}_1}{\mathrm{v}_3-\mathrm{v}_1+a_2(\mathrm{v}_3-\mathrm{v}_2)},
\end{equation*}
along with the density $\rho$, the fluid velocity $u$,  and the electric field $E$ as used in the single water-bag model. In this case, $\rho n_1$ simply corresponds to the density of the particles contained in the first bag of unit height. We see that $\int\rho n_1\ \mathrm{d} x$, i.e., the total number of particles in the first bag, is a Casimir invariant of Eq.~\eqref{eqBrackWB}, hence is a constant of motion. The contour velocities can be expressed explicitly with respect to these variables as follows: 
\begin{align*}
\mathrm{v}_1&=u+\frac{\rho}{2}\frac{a_2n_1(n_1-2)-1}{1+a_2},\\
\mathrm{v}_2&=u+\frac{\rho}{2}\frac{n_1(a_2n_1+2)-1}{1+a_2},\\
\mathrm{v}_3&=u+\frac{\rho}{2}\frac{a_2n_1^2+1}{1+a_2}.
\end{align*}
Expressed in terms of the variables $(\rho,u,n_1,E)$,  \eqref{eqBrackWB} takes the particularly simple form
\begin{multline}
\label{brackDouble}
\{F,G\}_2=\int\bigg[G_u\partial_x F_\rho-F_u\partial_x G_\rho+G_u\widetilde{F_E}-F_u\widetilde{G_E}\\
-\frac{1}{\rho}(F_1G_u-G_1F_u)\partial_x n_1+\frac{1+a_2}{a_{2}}\frac{F_1}{\rho}\partial_x\left(\frac{G_1}{\rho}\right)\bigg]\ \mathrm{d}x,
\end{multline}
where $F_1$ denotes the functional derivative of $F$ with respect to $n_1$. Bracket~\eqref{brackDouble} is closed which is, as in the single water-bag model, a property inherited from the original water-bag bracket given by Eq.~\eqref{eqBrackWB}. In terms of the variables $\rho$, $u$, $n_1$ and $E$, Hamiltonian~\eqref{hamWB} becomes
\begin{equation}
\label{hamDouble}
\mathcal{H}[\rho,u,n_1,E]=\frac{1}{2}\int\left[\rho u^2+\rho^3S_2(n_1)+E^2\right]\ \mathrm{d}x,
\end{equation}
where
\begin{equation}
\label{defS2}
S_2(n_1)=\frac{1+6a_2n_1^2+4a_2(a_2-1)n_1^3-3a_2^2n_1^4}{12(1+a_2)^2}.
\end{equation}
This shows that $S_2$ is a function of $n_1$ only. The specific internal energy now becomes  $U(\rho,n_1)=\rho^2S_2(n_1)/2$. As a consequence, in addition of the pressure defined as the thermodynamic conjugate variable of the density through the relation $P=\rho^2\partial U/\partial\rho$, we can define some potential $\mu_1$ as the conjugate variable of $n_1$,  such that $\mu_1=\rho^2S_2'(n_1)/2$. This shows that more accurate fluid models are obtained by introducing more information on the thermodynamic properties of the system through internal degrees of freedom. Bracket~(\ref{brackDouble}) and Hamiltonian~(\ref{hamDouble}) lead to Eqs.~(\ref{eqContinuity})-(\ref{eqAmpereRhoU}) and the  following additional equation:
\begin{equation*}
\partial_tn_1=\{n_1,\mathcal{H}\}_2=-u\partial_xn_1+ \frac{1+a_2}{ a_2\, \rho}\partial_x\left[\frac{\rho^2}{2}S_2'(n_1)\right].
\end{equation*}
The first term of this equation  is an advection term, while  the second is a flow term resulting from the potential $\mu_1=\rho^2S_2'(n_1)/2$. Indeed, analogously to the pressure $P$ that  drives a force $-\partial_xP$ in Eq.~\eqref{eqEuler}, here $\mu_1$ drives a flow $\partial_x\mu_1$.

Along with the Casimir invariants given by Eqs.~\eqref{Gauss2}, \eqref{meanE},  and \eqref{casRho}, the two water-bag model has the following global invariants:
\begin{align}
\label{casN1}
\overline{\rho n_1}&=\int\rho n_1\ \mathrm{d}x,\\
\label{casVdecale}
C_2&=\int \left(u+ \frac{a_2}{2(1+a_2)} \, \rho n_1^2\right) \mathrm{d}x.
\end{align}
The Casimir invariant given by Eq.~\eqref{casN1} is inherited from the original Vlasov-Amp\`ere model and amounts  to  conservation of the total entropy. Indeed, in statistical physics, the entropy of a system is related to its number of microstates. Here the microstates are given by the amount of particles in each bags. The invariant given by Eq.~\eqref{casVdecale} is a new conserved quantity and is generated by the reduction procedure. Noting that Eq.~\eqref{defS2} defines a bijection $g:n_1\in[0;1]\mapsto S_2=g(n_1)\in\mathbb{R}_+$ such that we can write $n_1=\kappa(S_2)$ where $\kappa=g^{-1}$, the previous invariants become 
\begin{align*}
\overline{\rho \kappa(S_2)}&=\int\rho \kappa(S_2)\ \mathrm{d}x,\\
C_2&=\int u+ \frac{a_2}{2(1+a_2)}\, \rho\kappa^2(S_2)\ \mathrm{d}x.
\end{align*}
Therefore,  we see that the two water-bag model is a particular  case of the more general three moments fluid model\cite{Perin14}. 

As stated previously, the two water-bag model is closed. Thus, we expect the associated fluid model to be closed too, and  the fourth reduced moment $S_3$   to be a function of $S_2$ only. In what follows, we choose $a_2\geq0$. This corresponds to a configuration in which the second bag is taller than the first, as depicted in Fig.~\ref{figS3}. The case $a_2<0$ is equivalent through the symmetry $v\rightarrow -v$ and $n_1\rightarrow 1-n_1$. By using the definition of the third order reduced moment given by Eq.~\eqref{eqS} with $i=3$,
we find
\begin{equation*}
S_3=-\frac{a_2(n_1-1)^2n_1^2(1+a_2n_1)^2}{4(1+a_2)^3}.
\end{equation*}
Since $S_3$ is a function of $n_1$ only and  $n_1=\kappa(S_2)$, we see  that $S_3$ is a function of $S_2$ only, i.e., $S_3=S_3(S_2)$ as expected. This relation was expected because it was  shown in Ref.~[\onlinecite{Perin14}] that it is the case for general  Hamiltonian closures  with three fluid moments obtained from the Vlasov equation. The fraction of particles in the first water-bag parametrizes the curve $n_1\mapsto[S_2(n_1),S_3(n_1)]$. This result corresponds to a closure for  the heat flux $q=\rho^4S_3/2$ as a function of the density $\rho$ and the pressure $P=\rho^3S_2$. We do not give the explicit relationship between $S_2$ and $S_3$ here because  it does not provide much information. However, the dependence of $S_3$ on $S_2$ is  plotted in Fig.~\ref{figS3} for different values of $a_2$.
We observe that as $a_2\rightarrow0$, we have $S_2\rightarrow1/12$ and $S_3\rightarrow0$. This is consistent with the results of Sec.~\ref{secSingle}. Indeed, for $a_2=0$ the two water-bags have the same heights and, as a consequence, merge into one such that we recover the values of the fluid moments corresponding to a single water-bag model. Moreover, as $a_2$ increases, the solution tends rapidly toward an equilibrium obtained with $a_2\rightarrow+\infty$.

In summary, the two water-bag model can be conveniently described by using appropriate fluid variables: the density $\rho$ and the fluid velocity $u$ are  natural fluid quantities that take into account, in particular through the definition of the kinetic energy $K=\int\rho u^2/2\ \mathrm{d}x$, the macroscopic energy of the system. The internal effects are described by the internal energy. Using the partitioning of the number of particles into two bags as the internal degree of freedom of the system, we can define the specific internal energy as $U(\rho,n_1)=\rho^2S_2(n_1)/2$. The two  water-bag model corresponds to a system with one internal degree of freedom described by $n_1$.

\subsection{Three water-bag  model}

\begin{figure}
\centering
\includegraphics[width=0.4\textwidth]{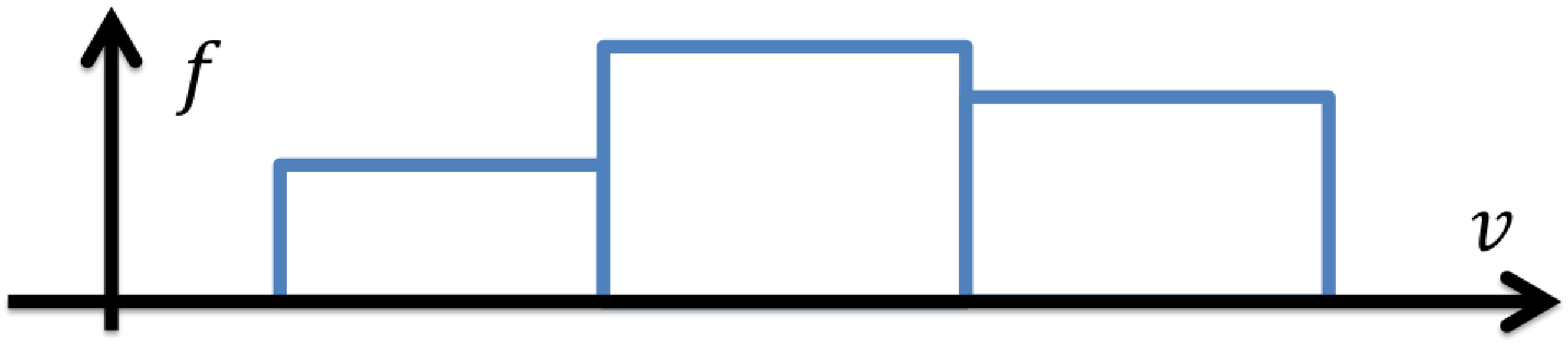}
\includegraphics[width=0.4\textwidth]{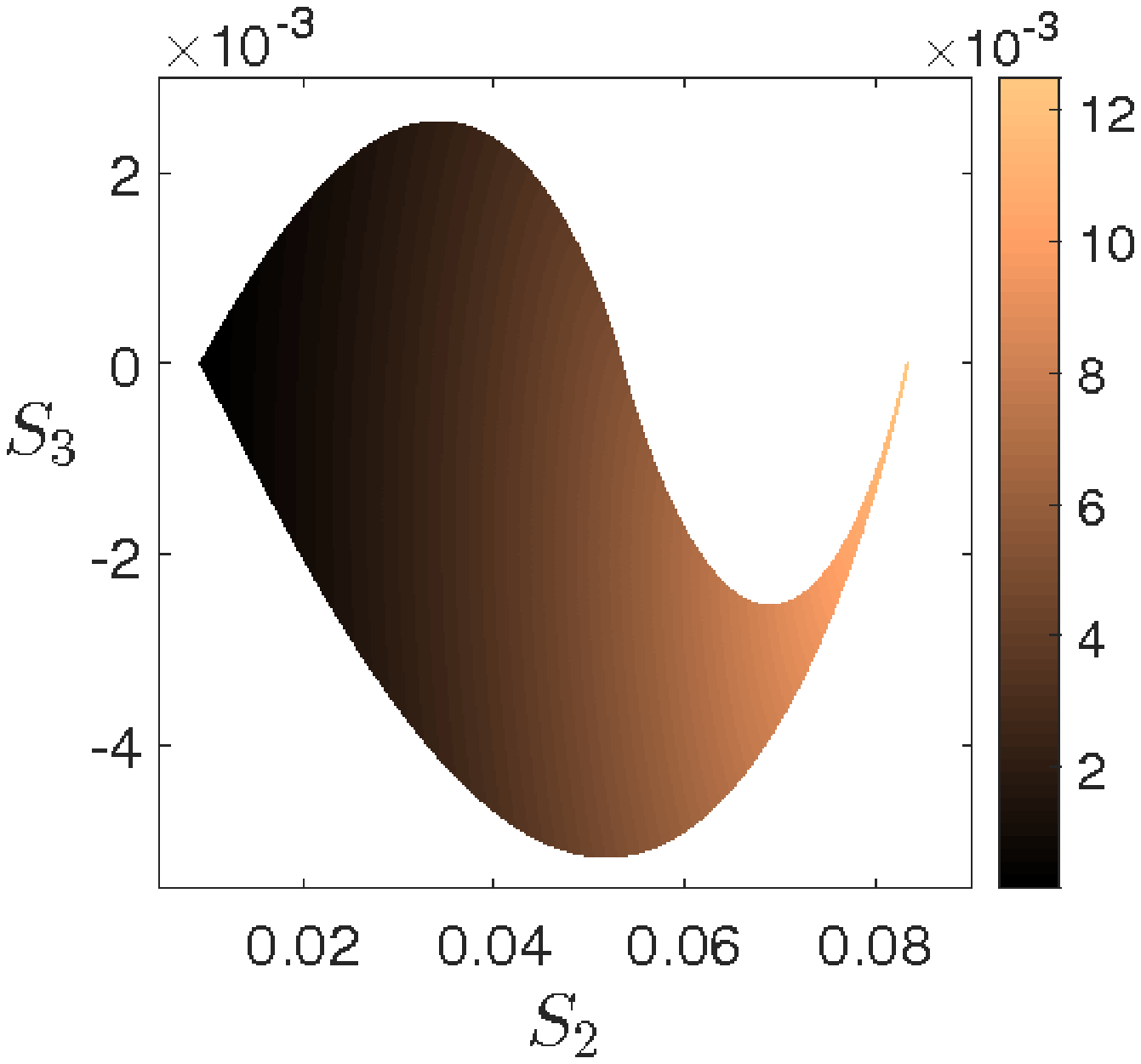}
\caption{Upper panel: sketch of a three water-bags distribution function whose typology exhibits a bell-shape. Lower panel: colormap of $S_4$ as a function of $S_2$ and $S_3$ for $(a_2,a_3)=(2,-1.75)$ corresponding to a distribution function given by the upper panel.}
\label{figS4Bump}
\end{figure}

\begin{figure}
\centering
\includegraphics[width=0.4\textwidth]{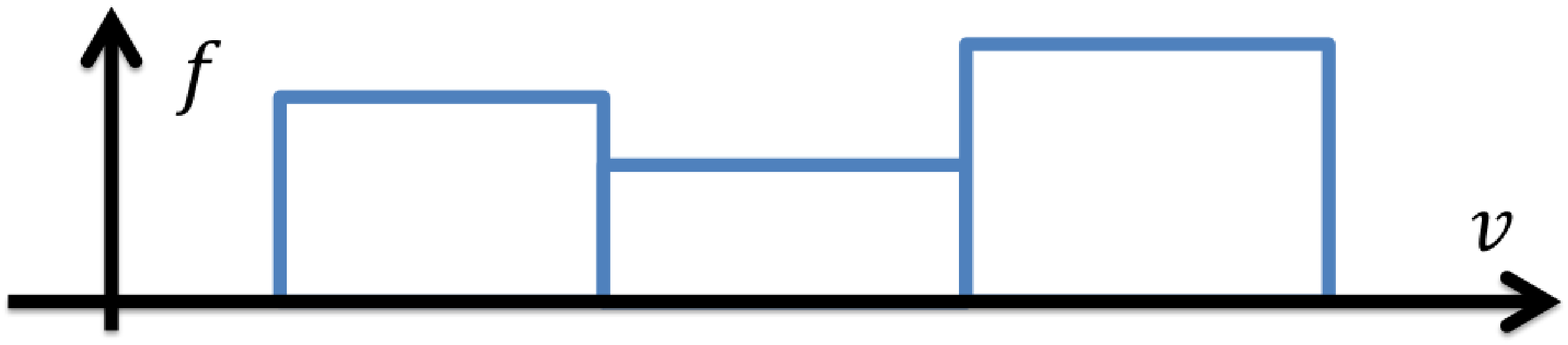}
\includegraphics[width=0.4\textwidth]{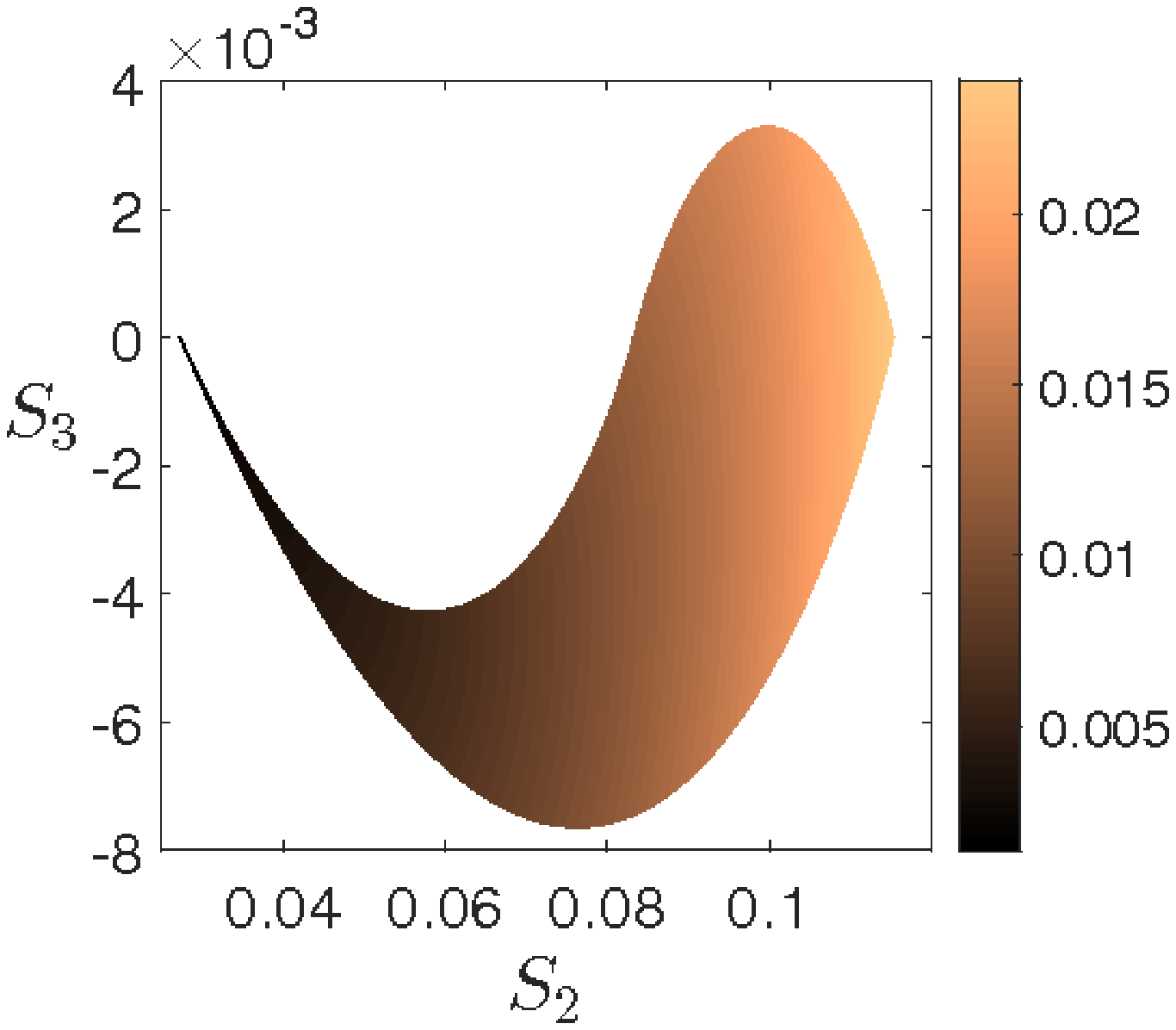}
\caption{Upper panel: sketch of a three water-bags distribution function whose typology exhibits a hole. Lower panel: colormap of $S_4$ as a function of $S_2$ and $S_3$ for $(a_2,a_3)=(-0.15,0.9)$ corresponding to a distribution function given by the upper panel.}
\label{figS4Hole}
\end{figure}

\begin{figure}
\centering
\includegraphics[width=0.4\textwidth]{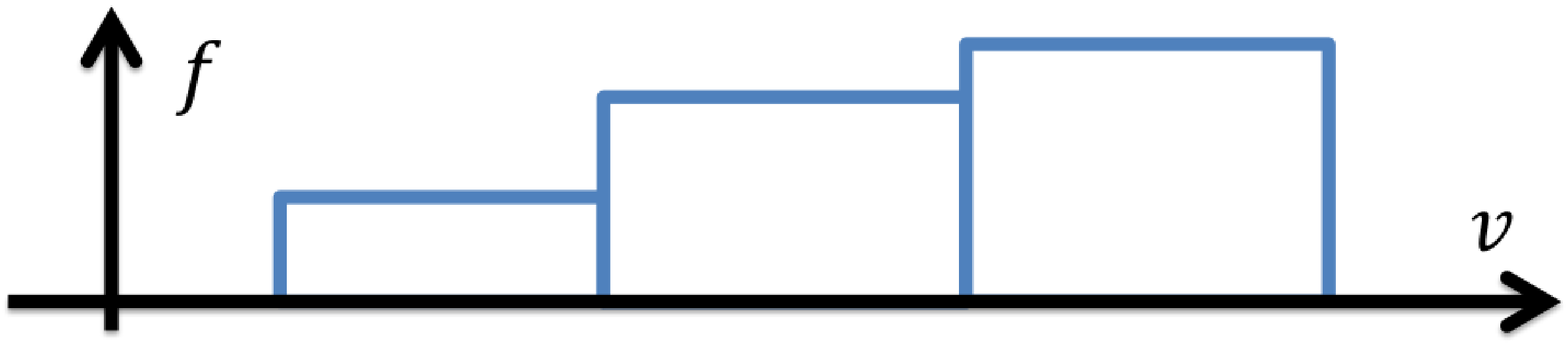}
\includegraphics[width=0.4\textwidth]{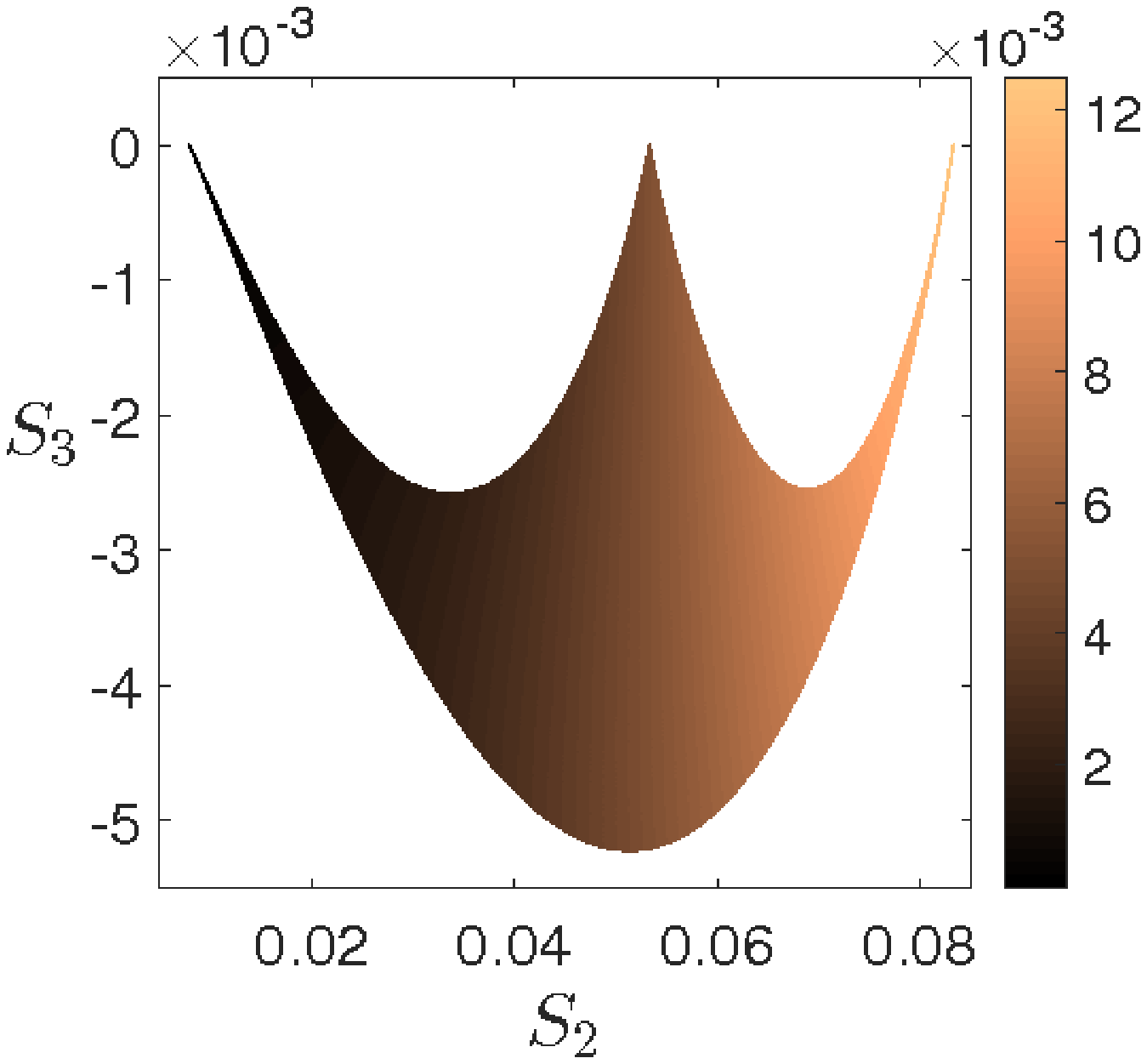}
\caption{Upper panel: sketch of a three water-bags distribution function whose typology exhibits a monotonic increase. Lower panel: colormap of $S_4$ as a function of $S_2$ and $S_3$ for $(a_2,a_3)=(0.25,2)$ corresponding to a distribution function given by the upper panel.}
\label{figS4Increasing}
\end{figure}

In this section, we demonstrate the usefulness of the new thermodynamical variables for linking the  water-bag and fluid models by considering the  three water-bag distribution function, whose expression is given by Eq.~\eqref{defWaterBag} for $N=3$. A three water-bag model is equivalent to a Hamiltonian fluid model with four fluid moments. Even though a particular closure based on dimensional analysis has been previously found for such fluid models\cite{Perin15}, there is currently no general Hamiltonian closure for fluid models with four moments. Finding all the closures is  difficult because it  requires solving non-linear PDEs obtained from the Jacobi identity.    We show here, by using the thermodynamic variables, that the water-bag distribution function with three water-bags is another, parameter-dependent closure for fluid models with four moments. This solution is simpler to compute than the general closure, and gives us some useful information. Different three water-bag distribution functions are depicted  in Figs.~\ref{figS4Bump}, \ref{figS4Hole} and \ref{figS4Increasing}.

Following the procedure of Sec.~\ref{secDouble}, we introduce the thermodynamic variable
\begin{equation*}
n_2=\frac{(1+a_2)(\mathrm{v}_{3}-\mathrm{v}_2)}{\mathrm{v}_4-\mathrm{v}_1+a_2(\mathrm{v}_4-\mathrm{v}_2)+a_3(\mathrm{v}_4-\mathrm{v}_3)},
\end{equation*}
in addition to $\rho$, $u$, $n_1$,  and $E$ used in the two water-bag model. Here $\rho n_2$ corresponds to the density of the particles contained in the second bag. The expressions of the contour velocities in terms of these variables are given by Eq.~\eqref{eqVk} and will not be detailed here. Expressed in terms of the variables $(\rho,u,n_1,n_2,E)$, Bracket~\eqref{eqBrackWB} takes the particularly simple form
\begin{multline}
\label{brackTriple}
\{F,G\}_3=\int\bigg[G_u\partial_x F_\rho-F_u\partial_x G_\rho+G_u\widetilde{F_E}-F_u\widetilde{G_E}\\
-\frac{1}{\rho}(F_iG_u-G_iF_u)\partial_x n_i+\beta_{ik}\frac{F_i}{\rho}\partial_x\left(\frac{G_k}{\rho}\right)\bigg]\ \mathrm{d}x,
\end{multline}
where $F_i$ denotes the functional derivative with respect to $n_i$ for $i\in\{1,2\}$ and where the summation over repeated indices from $1$ to $2$ is assumed. Here $\beta$ is a constant $2\times 2$ symmetric  matrix given by
\begin{equation*}
\beta=\frac{(1+a_2)}{a_2a_3}
\begin{pmatrix}
a_3 & -a_3\\
-a_3 & (1+a_2)(a_2+a_3)
\end{pmatrix}
.
\end{equation*}
In terms of the variables $\rho$, $u$, $n_1$, $n_2$, and $E$, Hamiltonian~\eqref{hamWB} becomes
\begin{equation}
\label{hamTriple}
\mathcal{H}[\rho,u,n_1,n_2,E]=\frac{1}{2}\int\left[\rho u^2+\rho^3S_2(n_1,n_2)+E^2\right]\ \mathrm{d}x,
\end{equation}
where $S_2(n_1,n_2)$ is given by Eq.~\eqref{Snu}. As in the two water-bag model, we can define the thermodynamic potential $\boldsymbol\mu=\partial U/\partial \mathbf{n}=\rho^2(\partial S_2/\partial \mathbf{n})/2$. Analogously to multi-components systems, the variables $n_i$ act as internal degrees of freedom that characterize the model through a partitioning of the particles with respect to their energy or, equivalently, their temperature. Indeed, it is  expected that a system like the  collisionless Vlasov-Amp\`ere system should  not thermalize and as a consequence is described by more than one temperature. Bracket~(\ref{brackTriple}) and Hamiltonian~(\ref{hamTriple}) lead to Eqs.~(\ref{eqContinuity}) - (\ref{eqAmpereRhoU}), and in addition the following equations:
\begin{equation*}
\partial_tn_i=\{n_i,\mathcal{H}\}_3=-u\partial_xn_i+\frac{1}{\rho}\beta_{ik}\partial_x\left(\frac{\rho^2}{2}\frac{\partial S_2}{\partial n_k}\right),
\end{equation*}
for $i\in\{1,2\}$,   where summation over repeated indices form 1 to 2 is assumed. These equations exhibit an advection term and a driving term through the existence of a potential $\boldsymbol\mu=\rho^2(\partial S_2/\partial \mathbf{n})/2$. 

Along with the Casimir invariants given by Eqs.~\eqref{Gauss2}, \eqref{meanE}, \eqref{casRho}, and \eqref{casN1}, Bracket~\eqref{brackTriple} has the following global invariants:
\begin{align*}
\overline{\rho n_2}&=\int\rho n_2\ \mathrm{d}x,\\
C_2&=\int \left\{u+\frac{\rho}{2}\left[\frac{a_2n_1^2}{(1+a_2)}+\frac{a_3(n_1+n_2)^2}{(1+a_2)(1+a_2+a_3)}\right]\right\}\ \mathrm{d}x\,.
\end{align*}
Thus, as for the other models, we have as many global Casimir invariants as dynamical variables. Moreover, there is some generalized velocity which is a common feature of all fluid models derived from Vlasov-Amp\`ere equations.

By using Eq.~\eqref{Snu}, we can express $S_2$, $S_3$ and $S_4$ as functions of $n_1$ and $n_2$. For any $(a_2,a_3)$, these functions define  a unique two-dimensional manifold $[S_2(n_1,n_2),S_3(n_1,n_2),S_4(n_1,n_2)]$.

There are mainly three configurations of interest for the distribution function, with  other typologies  obtained by using different symmetries. The first configuration  has  $0<-a_3<a_2$. This bell-shaped configuration is shown in Fig.~\ref{figS4Bump}. The second configuration has $0<-a_2<a_3$ and corresponds to a case in which the second bag is the smallest. Such a typology  exhibits a ``hole" in the distribution function shown in Fig.~\ref{figS4Hole}. The last typology has $a_2>0$ and $a_3>0$ and corresponds to a configuration with  the third bag taller than the second, which is taller than the first. Such a typology is shown in Fig.~\ref{figS4Increasing}. In  Figs.~\ref{figS4Bump}, \ref{figS4Hole}, and \ref{figS4Increasing}  we plot $S_4$ as a function of $S_2$ and $S_3$ for the  three configurations.

Observe,  despite the change in distribution function  typology, which result in a change in the typology of the manifolds defined by the closure, $S_4$ always increases as $S_2$ increases. Moreover, for a  configuration as depicted in  Fig.~\ref{figS4Increasing}, the sign of $S_3$ is fixed,  whereas it may vary for the other configurations depending on the respective widths of the water-bags. The same computation can  be performed for  higher order moments. In particular, a Hamiltonian fluid model for four moments  requires closures on the fourth and fifth order moments respectively, namely $S_4$ and $S_5$\cite{Perin15}.

\subsection{$N$ water-bag model}
\label{secNWB}

The method presented in the previous subsections can be extended to an arbitrary number of water-bags with a corresponding  arbitrary number of fluid moments. This is important because by increasing the number of water-bags we increase the accuracy of the description, allowing for more refined kinetic effects. This is consistent with the fact that the water-bag models  come from a discretization of the distribution function in velocity space. An analogy can be made with vibrations of structures\cite{Hammett93}. In these systems, the frequency spectrum is continuous. However, these models can be accurately described by a finite number of coupled springs with a discrete spectrum as long as the number of springs is sufficiently high. Thus by increasing the number of water-bags, yet keeping it finite, we can recover kinetic information about the system within the framework of a fluid description. Considering $N$ bags, we  define the following variables:
\begin{align*}
\rho&=-\sum\limits_{i=1}^{N+1}a_i\mathrm{v}_i,\\
u&=\frac{\sum\limits_{i=1}^{N+1}a_i\mathrm{v}_i^2}{2\sum\limits_{k=1}^{N+1}a_k\mathrm{v}_k},\\
n_l&=-\frac{(\mathrm{v}_{l+1}-\mathrm{v}_l)\sum\limits_{i=1}^la_i}{\sum\limits_{k=1}^{N+1}a_k\mathrm{v}_k},
\end{align*}
for all $1\leq l\leq N-1$. In this case, $\rho n_i$ corresponds to the density of the particles contained in the $i$-th bag for $1\leq i\leq N-1$. We see that $\int\rho n_i\ \mathrm{d} x$, i.e., the number of particles in the $i$-th bag, is a Casimir invariant of Eq.~\eqref{eqBrackWB}. The contour velocities can be expressed explicitly with respect to these variables such that
\begin{equation}
\label{eqVk}
\mathrm{v}_i=u+\rho\Psi_i
\end{equation}
for all $1\leq i\leq N+1$ and where
\begin{equation*}
\Psi_{N+1}=\frac{1}{2}\sum\limits_{i=1}^N\frac{a_i}{a_{N+1}^2}\left[1-\sum\limits_{k=1}^{N-1}n_k-a_{N+1}\sum\limits_{k=i}^{N-1}\frac{n_k}{A_k}\right]^2,
\end{equation*}
and
\begin{equation*}
\Psi_m=\Psi_{N+1}+\frac{1}{a_{N+1}}\left[1-\sum\limits_{k=1}^{N-1}n_k-a_{N+1}\sum\limits_{k=m}^{N-1}\frac{n_k}{A_k}\right],
\end{equation*}
for all $1\leq m\leq N$. Expressed in terms of the variables $(\rho,u,n_1,\dots,n_{N-1},E)$, Bracket~\eqref{eqBrackWB} takes the particularly simple form given by
\begin{multline}
\label{brackN}
\{F,G\}_N=\int\bigg[G_u\partial_x F_\rho-F_u\partial_x G_\rho+G_u\widetilde{F_E}-F_u\widetilde{G_E}\\
-\frac{1}{\rho}(F_iG_u-G_iF_u)\partial_x n_i+\beta_{ik}\frac{F_i}{\rho}\partial_x\left(\frac{G_k}{\rho}\right)\bigg]\ \mathrm{d}x,
\end{multline}
where $F_i$ denotes the functional derivative with respect to $n_i$ for $1\leq i\leq N-1$ and where the summation over repeated indices from $1$ to $N-1$ is again assumed. Here $\beta$ is a constant tridiagonal $(N-1)\times (N-1)$ symmetric matrix  given by
\begin{equation*}
\beta=
\begin{pmatrix}
\lambda_1 & -\lambda_1 & 0 & \dots & 0\\
-\lambda_1 & \lambda_1+\lambda_2 & -\lambda_2 & \dots & 0\\
0 & -\lambda_2 & \lambda_2+\lambda_3 & \dots & 0\\
\vdots & \vdots & \vdots & \ddots & \vdots\\
0 & 0 & 0 & \dots & \lambda_{N-1}
\end{pmatrix}
,
\end{equation*}
where
\begin{equation*}
\lambda_i=\frac{\sum\limits_{k=1}^ia_k\sum\limits_{l=1}^{i+1}a_l}{a_{i+1}}.
\end{equation*}
Bracket~\eqref{brackN} can be further simplified. Indeed, noting that $\beta$ is symmetric, hence diagonalizable, we introduce the variables
\begin{equation*}
\nu_i=\sum\limits_{k=1}^in_k,
\end{equation*}
for $1\leq i\leq N-1$. The quantity $\int\rho\nu_i\ \mathrm{d}x$ corresponds to the cumulative number of particles in the $i$ first bags. Eventually, Eq.~\eqref{brackN} takes the even simpler form
\begin{multline}
\label{brackEntropy2}
\{F,G\}_N=\int\bigg[G_u\partial_x F_\rho-F_u\partial_x G_\rho+G_u\widetilde{F_E}-F_u\widetilde{G_E}\\
-\frac{1}{\rho}(F_iG_u-G_iF_u)\partial_x \nu_i+\lambda_i\frac{F_i}{\rho}\partial_x\left(\frac{G_i}{\rho}\right)\bigg]\ \mathrm{d}x,
\end{multline}
where $F_i$ denotes the functional derivative of $F$ with respect to $\nu_i$ for $1\leq i\leq N-1$ and where summation over repeated indices from 1 to $N-1$ is assumed. The Hamiltonian associated with this model is given by
\begin{multline*}
\mathcal{H}[\rho,u,\nu_1,\dots,\nu_{N-1},E]\\
=\frac{1}{2}\int\left[\rho u^2+\rho^3S_2(\nu_1,\dots,\nu_{N-1})+E^2\right]\ \mathrm{d}x,
\end{multline*}
where the reduced moments can be computed from Eq.~\eqref{eqS} and are given by
\begin{equation*}
S_i(x,t)=-\frac{1}{(i+1)\rho^{i+1}}\sum\limits_{k=1}^{N+1}a_k[\mathrm{v}_k(x,t)-u(x,t)]^{i+1},
\end{equation*}
for all $i\geq2$. By using Eq.~\eqref{eqVk}, this eventually becomes 
\begin{equation}
\label{Snu}
S_i=\frac{-1}{(i+1)}\sum\limits_{k=1}^{N+1}a_k\xi_k^{i+1}(\nu_1,\dots,\nu_{N-1}),
\end{equation}
where
\begin{align*}
&\xi_{N+1}=\frac{1}{2}\sum\limits_{i=1}^N\frac{a_i}{a_{N+1}^2}\left[1-\nu_{N-1}-a_{N+1}\sum\limits_{k=i}^{N-1}\frac{\nu_k-\nu_{k-1}}{A_k}\right]^2,\\
&\xi_m=\xi_{N+1}+\frac{1}{a_{N+1}}\left[1-\nu_{N-1}-a_{N+1}\sum\limits_{k=m}^{N-1}\frac{\nu_k-\nu_{k-1}}{A_k}\right],
\end{align*}
for all $m$ such that $2\leq m\leq N-1$. This shows that the reduced moments $S_i$ are functions of the thermodynamic variables $\nu_i$ only. The equations of motion of the system are given by Eqs.~(\ref{eqContinuity})-(\ref{eqAmpereRhoU}) and
\begin{equation*}
\partial_t \nu_i=\{\nu_i,\mathcal{H}\}_N=-u\partial_x \nu_i+\frac{\lambda_{i}}{\rho}\partial_x\left(\frac{\rho^2}{2}\frac{\partial S_2}{\partial\nu_i}\right),
\end{equation*}
for all $1\leq i\leq N-1$. Along with the Casimir invariants given by Eqs.~\eqref{Gauss2}, \eqref{meanE}, and \eqref{casRho}, the $N$ water-bag model has the following global invariants:
\begin{align*}
\overline{\rho \nu_i}&=\int\rho \nu_i\ \mathrm{d}x,\\
C_2&=\int\bigg(u +\frac{\rho}{2}\sum\limits_{k=1}^N\frac{a_k}{a_{N+1}^2}\Big[\nu_{N-1}
\\
&\hspace{1.5cm} +a_{N+1}\sum\limits_{l=k}^{N-1}\frac{\nu_l-\nu_{l-1}}{A_l}\Big]^2\bigg)\ 
\mathrm{d}x.
\end{align*}
for all $1\leq i\leq N-1$.

\begin{figure}
\centering
\includegraphics[width=0.4\textwidth]{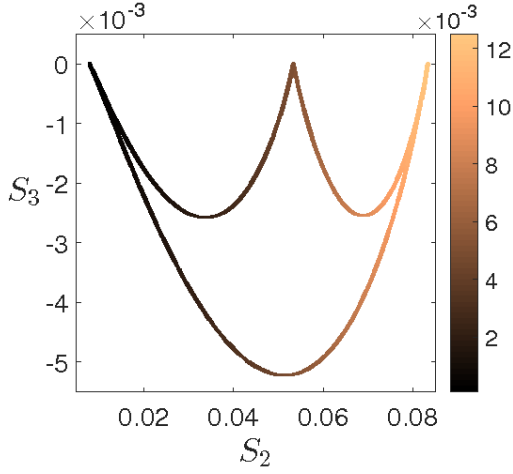}
\caption{Closures of the two water-bags model that define the projection of the edges of the closure of the three water-bags model.}
\label{figRand}
\end{figure}

The closure provided by the $N$ water-bag model is not straightforward. Indeed, the closure is such that $S_{N+1}=S_{N+1}(S_2,\dots,S_N)$, which defines an  $N-1$-dimensional manifold in $\mathbb{R}^N$ parametrized by $(\nu_1,\dots,\nu_{N-1})$. However, one can apply the tools developed throughout the previous sections to visualize such a manifold. Indeed, consider, e.g., a three water-bag  distribution function as shown in  Fig.~\ref{figS4Increasing}  and consider all the distribution functions (three in this example) obtained  by combining two of these bags.  These are shown  in Fig.~\ref{figRand}.
We see that the edges of the manifold defined by the four fluid moment closure $S_4=S_4(S_2,S_3)$ correspond to the closures of the three fluid moments models associated with every combination of two water-bags of the initial three water-bags distribution function. Analogously, the projections on the $(S_2,S_3,S4)$ space of the edges of the manifold defined by the closure $S_{N+1}=S_{N+1}(S_2,\dots,S_N)$ in the $(S_2,S_3,\dots,S_{N+1})$ space correspond to the closure of the three fluid moment models associated with every combination of two water-bags of the initial $N$ water-bag distribution function. 

\begin{figure}
\centering
\includegraphics[width=0.4\textwidth]{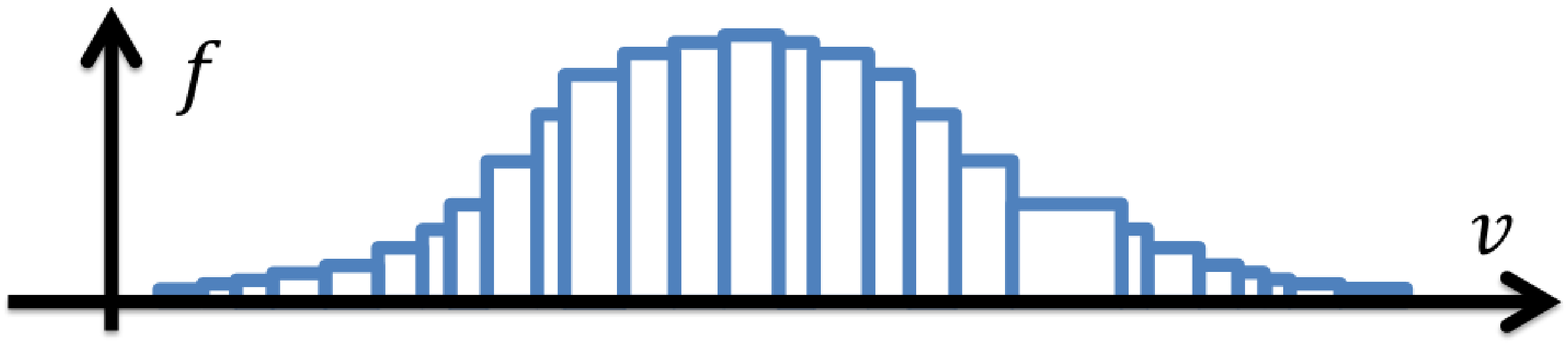}
\includegraphics[width=0.4\textwidth]{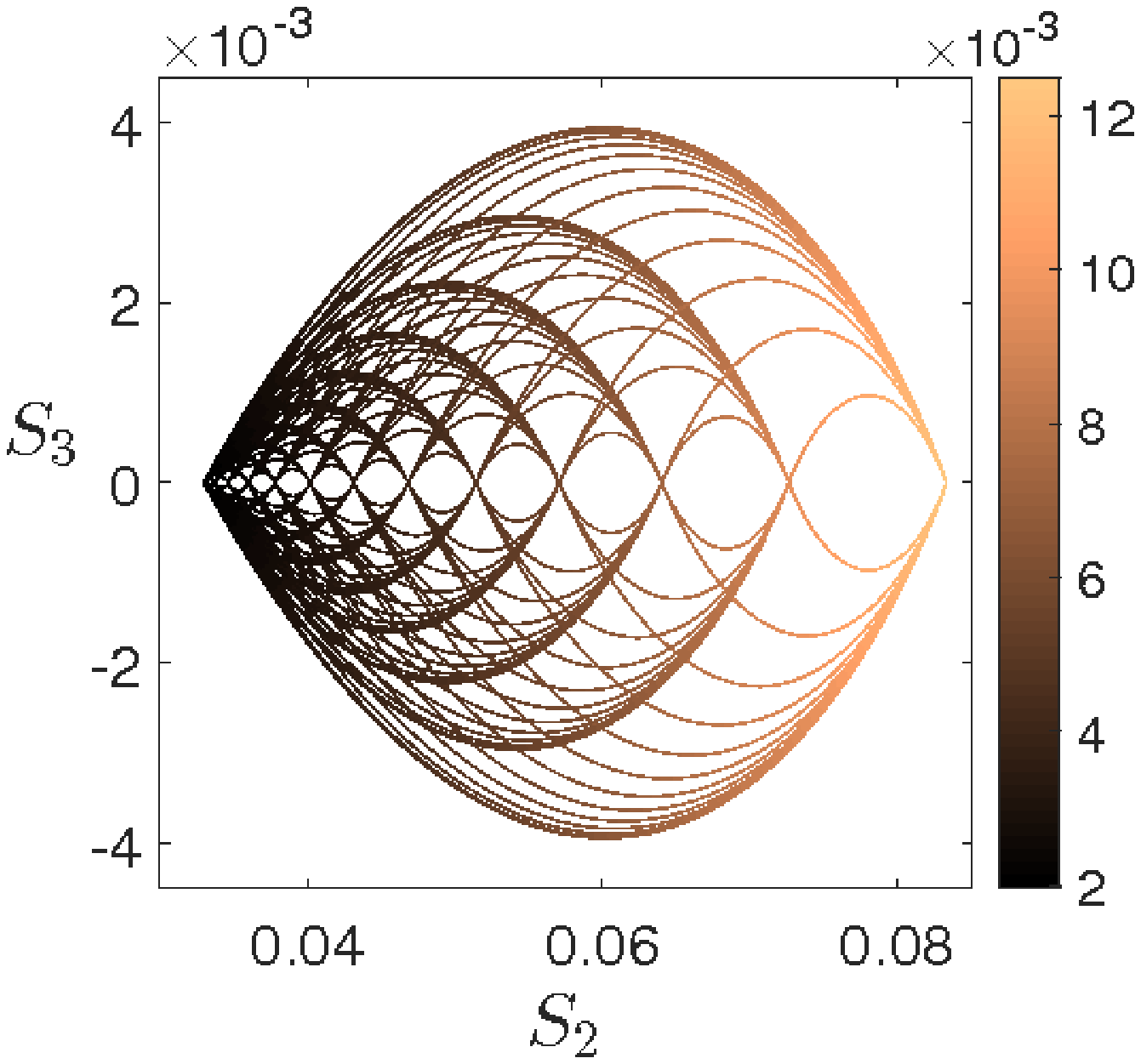}
\caption{Upper panel: sketch of a bell-shaped water-bag distribution function with twenty-seven bags. Lower panel: projection of the edges of the manifold defining the closure for the distribution function given by the upper panel.}
\label{figGaussienne}
\end{figure}

We illustrate the above edge description  with the following example. Consider a Maxwellian distribution approximated with a twenty-seven water-bag distribution function as  shown in Fig.~\ref{figGaussienne}. The corresponding fluid closure is such that $S_{28}=S_{28}(S_2,\dots,S_{27})$\cite{Perin15}. The projection of this high dimensional manifold on the $(S_2,S_3,S_4)$ space is depicted in Fig.~\ref{figGaussienne}.
Consequently, we see that the information about the whole system is given by all the possible couplings between two different water-bags. This makes the study of systems with a high number of fields easier as it eventually reduces to the study of coupled subsystems with three fields.

Inserting arbitrary functions in front of the terms $(F_i/\rho)\partial_x(G_i/\rho)$ in  \eqref{brackEntropy2} may allow us to extend this bracket to more general Poisson brackets of the form
\begin{multline}
\label{brackExtent}
\{F,G\}=\int\bigg\{G_u\partial_x F_\rho-F_u\partial_x C_\rho+G_u\widetilde{F_E}\\
-F_u\widetilde{G_E}-\frac{1}{\rho}(F_iG_u-G_iF_u)\partial_x \nu_i\\
+\sigma_i(\nu_i)\left[\frac{F_i}{\rho}\partial_x\left(\frac{G_i}{\rho}\right)-\frac{G_i}{\rho}\partial_x \left(\frac{F_i}{\rho}\right)\right]\bigg\}\ \mathrm{d}x,
\end{multline}
where the $\sigma_i$ are arbitrary functions. In addition, we consider Hamiltonians of the general form
\begin{equation*}
\mathcal{H}=\frac{1}{2}\int\left[\rho u^2+\rho^3S_2(\nu_1,\dots,\nu_{N-1})+E^2\right]\ \mathrm{d}x,
\end{equation*}
where now the dependence of $S_2$ on $\nu_i$ is arbitrary. The choice of the unknown functions $\sigma_i$ and the reduced moment $S_2(\nu_1,\dots,\nu_{N-1})$ should be based on physical arguments. As an example, we expect homogeneous initial conditions (i.e., $\rho=\rho_0$, $u=0$ and $\nu_i=\nu_{i0}$) to be linearly stable. This leads to constraints on $\sigma_i$ and $S_2(\nu_1,\dots,\nu_{N-1})$. It is possible show, e.g., that the two water-bag closure is always stable with respect to a homogeneous equilibrium. Along with the ones given by Eqs.~\eqref{casRho} and \eqref{meanE}, the global Casimir invariants of the extended water-bag bracket given by Eq.~\eqref{brackExtent} are
\begin{align*}
\overline{\rho\kappa_l}&=\int\rho\kappa_l\ \mathrm{d}x,\\
C_2&=\int\left(u+\frac{\rho}{4}\sum\limits_{l=1}^{N-1}\kappa^2_l\right)\ \mathrm{d}x,
\end{align*}
where $\kappa_l'=1/\sqrt{\sigma_l}$ for all $1\leq l\leq N-1$.\cite{foot4} We also notice that Brackets~\eqref{brackExtent} for $N=2$ are the most general Poisson brackets with three moments\cite{Perin14}. Whether or not this is the case for any $N$ is an open question.

\section{Summary}

In summary, we exhibited a method for constructing   Hamiltonian fluid models with an  arbitrary number of fluid moments  from the Vlasov-Amp\`ere system. This construction relies on the  Hamiltonian structure of the water-bag representation of a distribution function. We introduced  suitable fluid variables, based on thermodynamic considerations,  to replace the less meaningful contour velocities. The density and the fluid velocity were used to describe macroscopic phenomena, while the partitioning of the particles into the different bags was used to define  internal degrees of freedom in the system,  accounting  for microscopic effects. By using these variables, we were able to link the water-bag and fluid models and to make explicit the corresponding closures. We showed that, for an arbitrary number of water-bags, the general associated closure can be constructed from  knowledge of the couplings between all the other water-bags. Based on these results, we proposed a general $N$ field fluid model to describe plasmas with $N-2$ internal degrees of freedom.

\section*{Acknowledgments}

This work was supported by the Agence Nationale de la Recherche (ANR GYPSI). PJM was supported by U.S. Department of Energy Contract \#DEFG02-04ER54742.


\begin{thebibliography}{10}
\bibitem{foot1}Numerous works in the literature use the following alternative definition for the water-bag distribution\citep{Roberts67,Bertrand68,Berk70,Morel07,Gravier08,Morel08,Besse09,Gravier13}:
\begin{equation*}
\tilde{f}_N(x,v,t)=\sum\limits_{i=1}^Nb_i\left(\Theta[v-\mathrm{v}_i^-(x,t)]-\Theta[v-\mathrm{v}_i^+(x,t)]\right),
\nonumber
\end{equation*}
where the heights $b_i$ are constants. As a consequence, there are $2N$ fields (resp. $N$ parameters), namely $\mathrm{v}_i^+(x,t)$ and $\mathrm{v}_i^-(x,t)$ (resp. $b_i$) for all $1\leq i\leq N$. By using our notation, this formulation can be written $\tilde{f}_N(x,v,t)=f_{2N-1}(x,v,t)$ where $a_i=-a_{2N+1-i}=b_i$, $\mathrm{v}_i(x,t)=\mathrm{v}_i^-(x,t)$ and $\mathrm{v}_{2N+1-i}(x,t)=\mathrm{v}_i^+(x,t)$ for all $1\leq i\leq N$. Consequently, the formulation given by this representation only has an even number of contour velocities $\mathrm{v}_i(x,t)$ (field variables) and  only  has  $N$ independent free parameters. This corresponds to $N$ adiabatic fluids.
\bibitem{foot4}  One can see that there is a singularity for $\sigma_i=0$. This can be interpreted as a phase transition. Indeed, for $\sigma_i=0$, the corresponding field $\nu_i$ would only be advected by the velocity field $u$, hence acting as a passive tracer which would not influence the dynamics \cite{Morrison98}. This removes a degree of freedom in the definition of the internal energy.
\end{thebibliography}
\end{document}